\documentclass{aipproc}
\layoutstyle{8x11single}

\begin{document}

\title[Simulating Large-Format Arrays]{Simulating the Performance of 
Large-Format Sub-mm Focal-Plane Arrays}
\author{E. L. Chapin\footnotemark\footnotetext{{\bf e-mail:} {\em
echapin@inaoep.mx}}~}{
  address={Instituto Nacional de Astrof\'isica, \'Optica y Electr\'onica,
  Apartado Postal 51 y 216, 72000, Puebla, Mexico}}
\author{D. H. Hughes}{
  address={Instituto Nacional de Astrof\'isica, \'Optica y Electr\'onica,
  Apartado Postal 51 y 216, 72000, Puebla, Mexico} }
\author{B. D. Kelly}{
  address={UK Astronomy Technology Centre, Blackford Hill, EH9 3HJ, 
Edinburgh, UK} }
\author{W. S. Holland}{
  address={UK Astronomy Technology Centre, Blackford Hill, EH9 3HJ, 
Edinburgh, UK} }

\copyrightyear{2001}

\begin{abstract}
A robust measurement of the clustering amplitude of the sub-mm population of
starburst galaxies requires large-area surveys ($\gg 1$ deg$^2$). The
largest-format arrays subtend only 10 arcmin$^2$ on the sky and hence
scan-mapping is a necessary observing mode.  Providing realistic
representations of the extragalactic sky and atmosphere, as the input to a
detailed simulator of the telescope and instrument performance, allows
important decisions to be made about the design of large-area fully-sampled
surveys and observing strategies. In this paper we present preliminary
simulations that include detector noise, time-constants and array geometry,
telescope pointing errors, scan speeds and scanning angles, sky noise and sky
rotation.
\end{abstract}

\date{\today}
\maketitle

\section{Generating Realistic Synthetic Time-Series}
A mapping simulator has been developed to generate synthetic bolometer
time-series data and associated astrometric information which are then run
through a realistic reduction pipeline ({\em
http://www.inaoep.mx/$\sim$echapin/scansim.html}). This process allows 
one to develop and test the observing strategy and analysis software in
advance of instrument delivery.  Furthermore it is possible to assess the
impact of instrumental design aspects on the science objectives.

 \begin{figure}[hbtp]
\resizebox{0.45\textwidth}{!}{\includegraphics{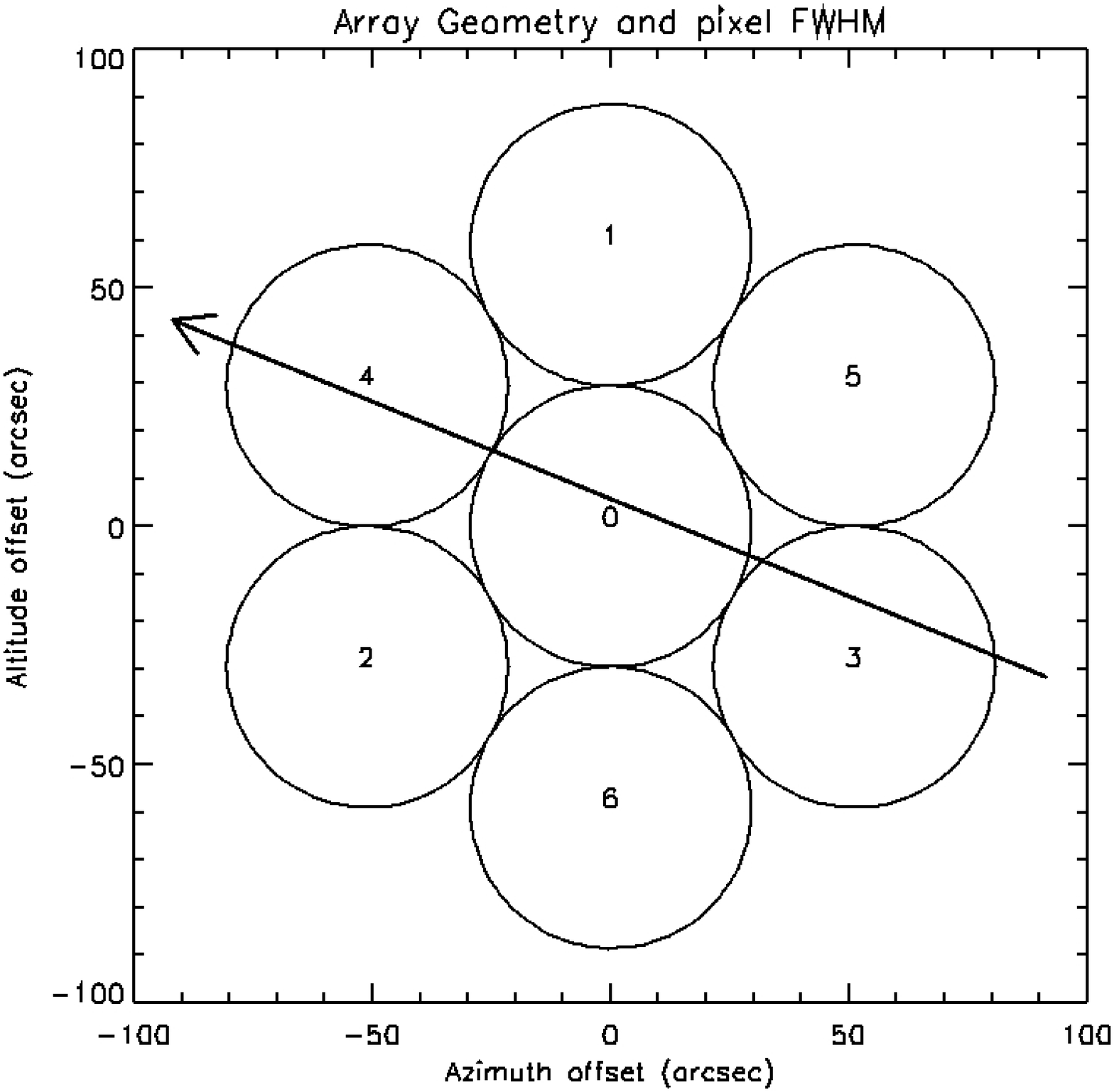}}     
\resizebox{0.45\textwidth}{!}{\includegraphics{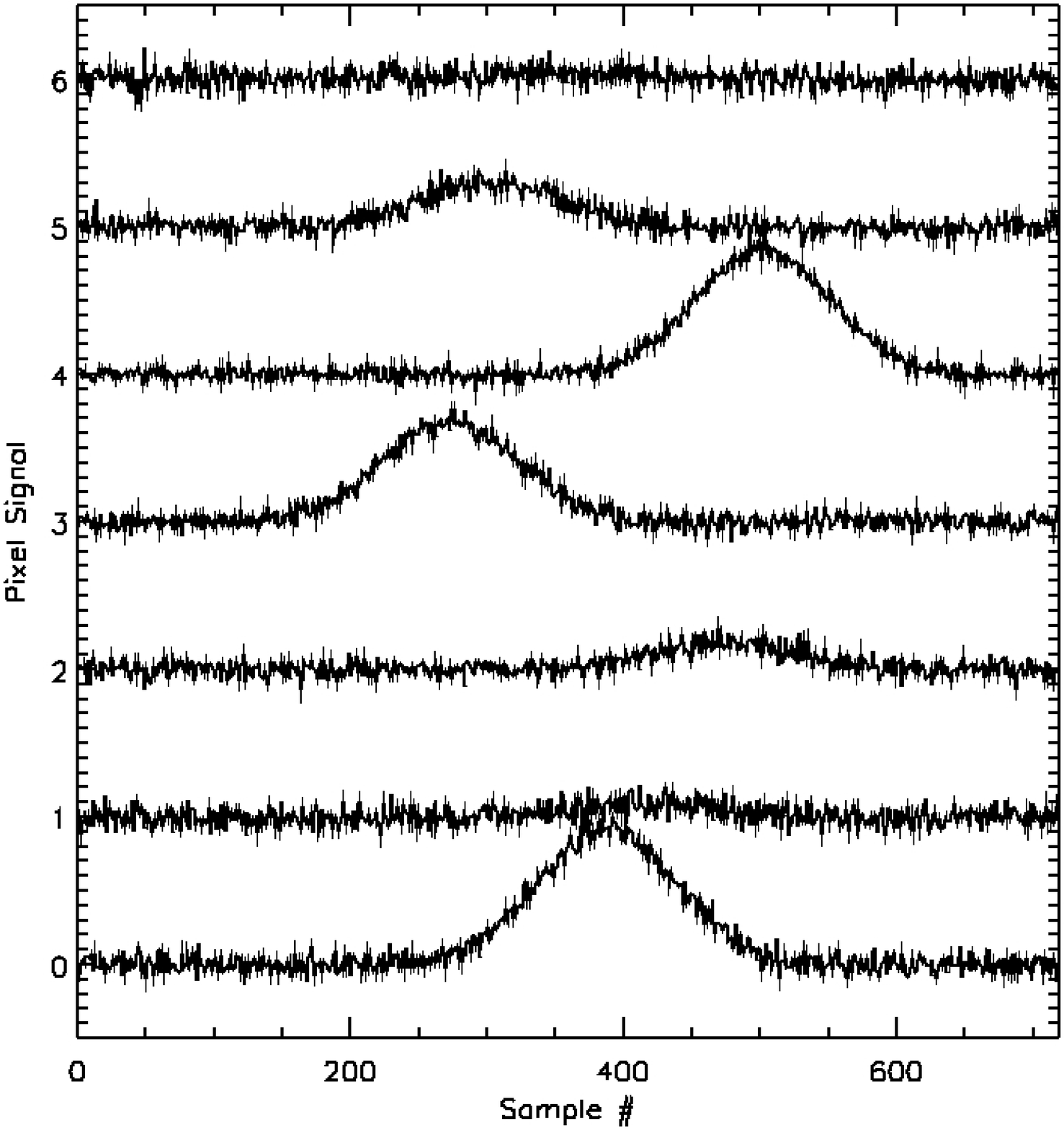}}    
\caption{7-element, 500$\mu$m 1-$F\lambda$ spaced hexagonal
close-packed array geometry with 59$''$ FWHM beams, and bolometer signals for
a scan across a point source (path of source indicated by arrow).}
\label{fig:sim}
\end{figure}

The simulated sky, typically generated by convolving a synthetic
catalogue of extra-galactic sources with the telescope beam
\cite{simulations}, is assigned equatorial coordinates on the celestial
sphere. The geometry of the array is defined and the beam positions on the sky
are determined relative to the telescope bore-sight. The telescope is then
arbitrarily located and the beams are scanned across the simulated sky model
(in some pattern on an Alt-Az coordinate system for a given local time) and,
using a rigorous calculation of the astrometry, the positional information and
{\em noiseless} flux density time-series is determined for each bolometer in
the array.

More realistic time-series are produced by incorporating features
characteristic of real instruments, the pointing performance of the telescope,
and the non-negligible contribution of the atmosphere (sky noise and
attenuation).

Sources of instrumental noise in the time-series include: 1) addition of an
uncorrelated Gaussian system noise component; 2) multiplication by gain
factors (or different responsivities) for each detector; 3) addition of
independent low frequency $1/f$ noise for each bolometer time-series to model
the slowly-changing detector baselines in long duration scans; 4) convolution
with an $e^{-t/\tau}$ impulse function to mimic the instantaneous response of
the detectors. The instrumental noise components for a linear detector are
summarized by the following expression for the response:

\begin{equation}
I_{measured} = G \times F + I_{dark} \label{response}
\end{equation}

\noindent where $G$ is the detector gain, $F$ is the incident flux and 
$I_{dark}$ is the {\em dark current} consisting of both broad-band and 
$1/f$ noise components. The {\em perfect} astrometry may also be 
{\em corrupted} by adding random and systematic pointing errors. 
Fig.~\ref{fig:sim} shows the time-series for a 7-pixel feed-horn coupled 
array scanned across a point source at 500$\mu$m.

\section{Sky Noise}

\begin{figure}[hbtp]
\resizebox{0.9\linewidth}{!}{\includegraphics{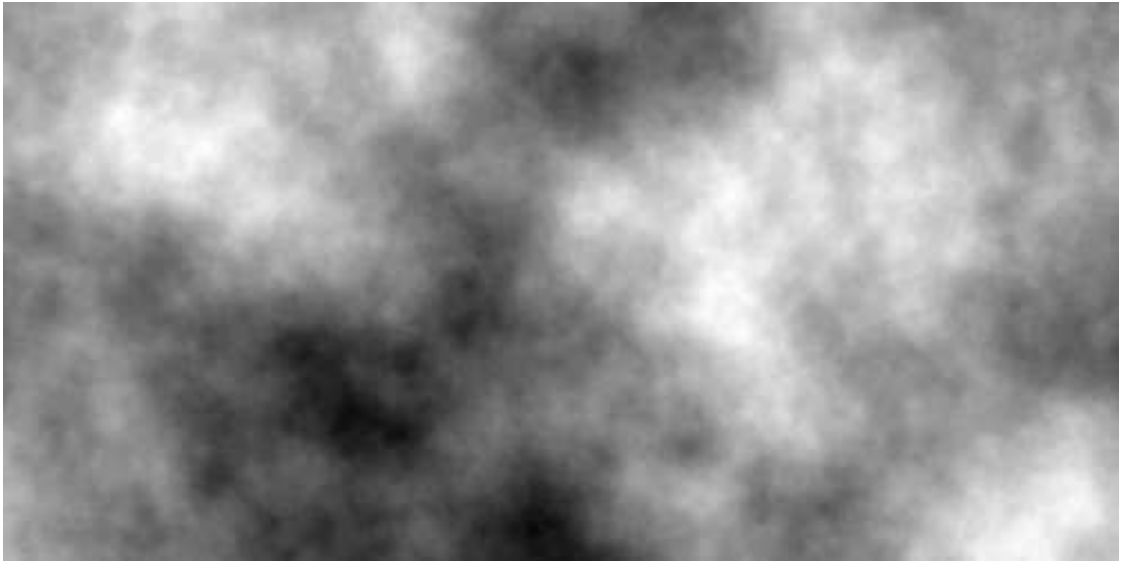}}
\caption{A 360 $\times$ 180 degree$^2$ simulation of sub-mm sky emission
at 850$\mu$m from cells of water vapour distributed in a flat plane at an
altitude of 600m above the telescope.}
\label{fig:sky}
\end{figure}

\begin{figure}[hbtp]
\resizebox{0.5\textwidth}{!}{\includegraphics{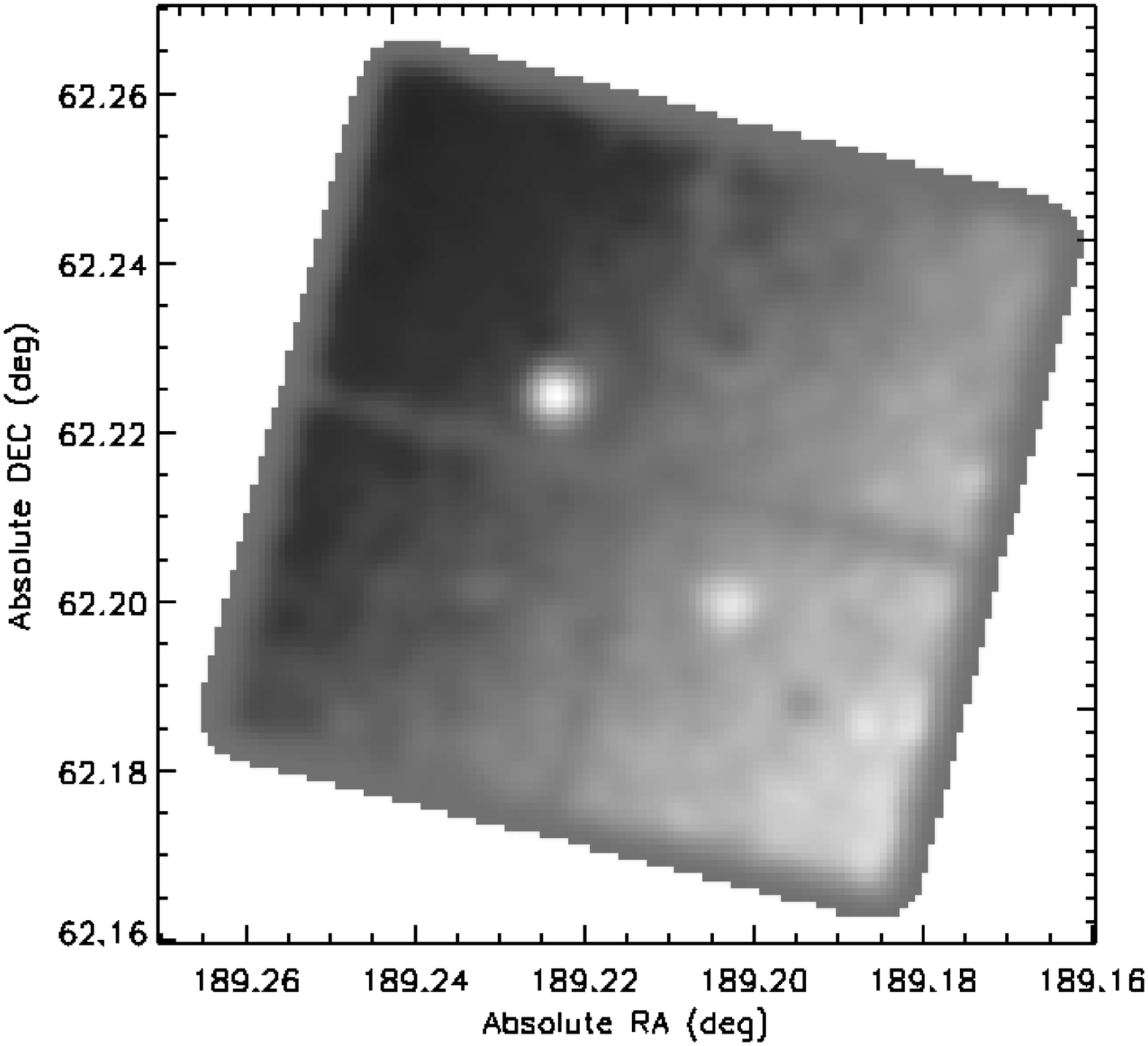}}
\resizebox{0.5\textwidth}{!}{\includegraphics{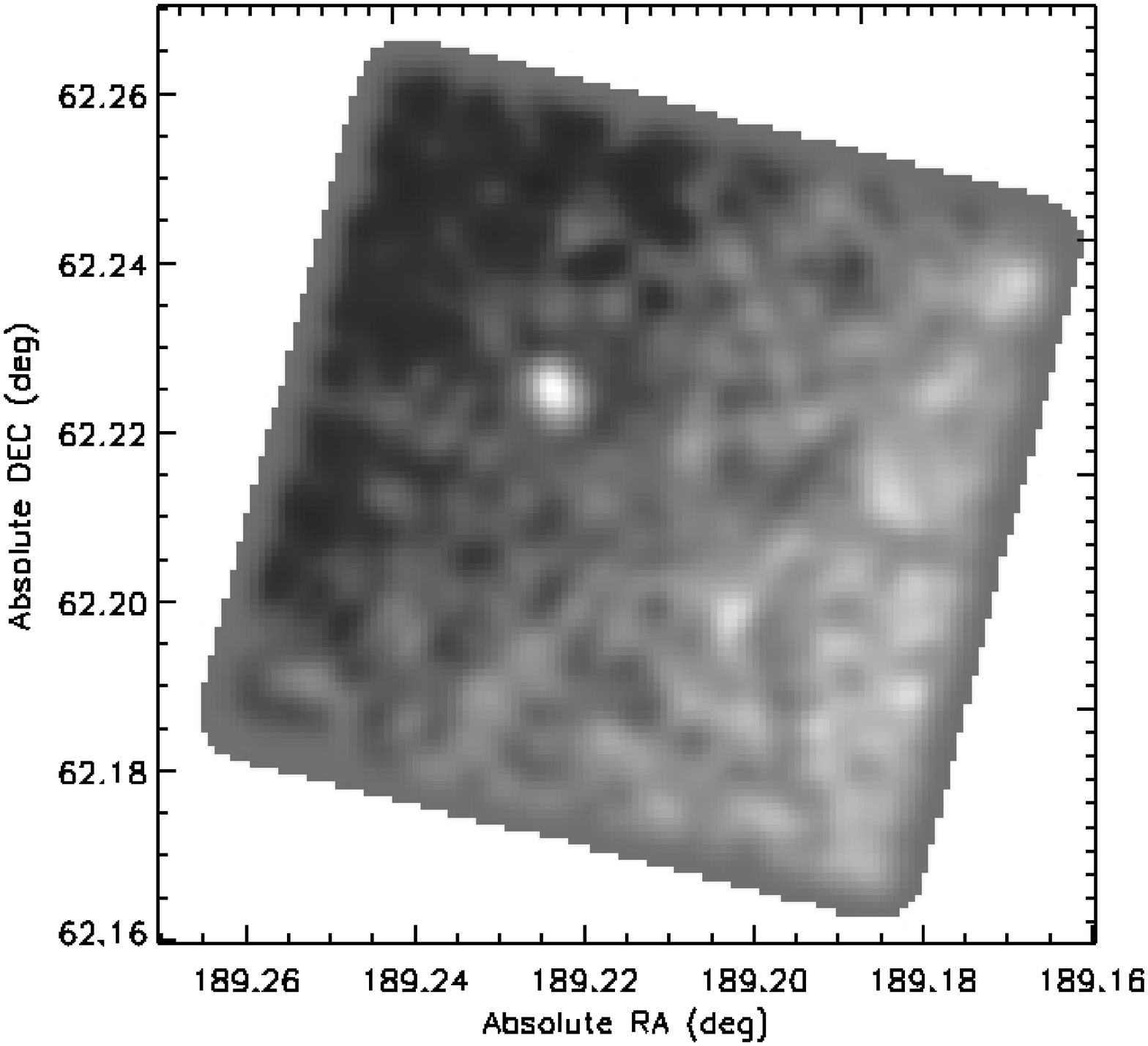}}
\caption{Simulations of one-second point-and-stare total power
observations with a fully-sampled (0.5-$F\lambda$) $40 \times 40$
pixel monolithic array at 850$\mu$m with 14$''$ FWHM beams.  In the case
of perfect flat-fielding (left panel), two significant sources with a
S/N of 25 and 12$\sigma$ are clearly observed against a typical
sky-background gradient. In contrast, the fainter of the sources
blends into the background noise when the gains of the bolometers are only
known to a precision of $\sim 1\%$ (right panel).}
\label{fig:point}
\end{figure}

Sky noise at sub-mm wavelengths is due to variable emission from cells
of water vapour moving across the field-of-view of the telescope. This
sky signal is significantly greater than the astronomical signal, and
errors introduced through its imperfect subtraction lead to increased
noise and artifacts in the reduced data. Thus, this important 
component of noise must be considered in all realistic simulations of
ground-based observations. 

Data from the SCUBA camera \cite{scuba} on the James Clerk Maxwell Telescope on Mauna Kea
suggests the sky generates $1/f$-like noise that is correlated across the
array. An empirical model that reproduces the observed spectrum of signals
measured with the SCUBA array at 850$\mu$m assumes the sky consists of cells
of water vapour, generated by convolving Gaussian noise with a symmetrical
$(1/f)^{11/6}$ function, where $f$ is the 2-D spatial frequency, distributed
in a flat-plane at some fiducial altitude, moving at the wind-speed above the
telescope aperture (Fig.~\ref{fig:sky}).  This planar sky-model is passed
across the simulated beams to produce a strong, variable sky signal.

Each 850$\mu$m pixel in the SCUBA2 array under good weather conditions will be
illuminated by a sky background power of 8pW with a photon noise in a 1s
integration of $4.3 \times 10^{-5}$pW. While the total power of the sky may be
removed by subtracting the median signal of the off-source sky bolometers,
strong gradients still remain and can dominate the astronomical signals. DC
sky measurements with SCUBA are used to estimate these gradients by
transforming the time-series into a spatial signal via the recorded wind
speed, assuming the emission is from a fixed altitude above the telescope, and
scaling it accordingly for the different background loading of the SCUBA2
pixels. The estimated median change in the sky level across an 8 arcmin field
of view is $7 \times 10^{-4}$pW (about an order-of-magnitude greater than the
photon noise). These estimates of the expected gradients and DC power levels
allow the proper scaling and bias of the planar-cloud model used to generate
the sky component of bolometer signals in the simulation.

Effective subtraction of the sky in the case of total power
measurements depends critically upon the knowledge of detector gains
and drifts. In Equation \eqref{response}, $G$ and $I_{dark}$ vary from
pixel to pixel and need to be removed using separate operations. The
gain $G$ is slowing-changing, and characterized by performing
flat-field observations with long integrations. The flat-fielding error 
must be smaller than the photon noise measured in
the longest astronomical integration. In addition $G$ is multiplied
by the large sky signal ($\gg$ astronomical signal), small errors in
its measurement will lead to residual artifacts upon dividing it out
of the response. An initial estimate suggests that $G$ will need to be
measured to an accuracy of $\ll 1\%$. $I_{dark}$ consists of
broad-band (white) noise plus a slowly varying $1/f$ component; the
drift may be subtracted by measuring ``dark'' frames (1s integrations)
in the absence of sky signal every few minutes.

Fig.~\ref{fig:point} demonstrates, for pointed total-power observations, two
examples of the effects described above. In conclusion, the requirement to
flat-field at sub-mm wavelengths with a precision of $\ll 1\%$ presents a
major obstacle to the efficient use of large-format monolithic arrays on
telescopes operated in total power mode.

\end{document}